\documentstyle[twoside]{article}

\catcode`\@=11
\long\def\@makefntext#1{
\protect\noindent \hbox to 3.2pt {\hskip-.9pt  
$^{{\eightrm\@thefnmark}}$\hfil}#1\hfill}		

\def\thefootnote{\fnsymbol{footnote}}
\def\@makefnmark{\hbox to 0pt{$^{\@thefnmark}$\hss}}	
	
\def\ps@myheadings{\let\@mkboth\@gobbletwo
\def\@oddhead{\hbox{}
\rightmark\hfil\eightrm\thepage}   
\def\@oddfoot{}\def\@evenhead{\eightrm\thepage\hfil
\leftmark\hbox{}}\def\@evenfoot{}
\def\sectionmark##1{}\def\subsectionmark##1{}}

\oddsidemargin=\evensidemargin
\renewcommand{\thefootnote}{\fnsymbol{footnote}}

\newcounter{sectionc}\newcounter{subsectionc}\newcounter{subsubsectionc}
\renewcommand{\section}[1] {\vspace{12pt}\addtocounter{sectionc}{1} 
\setcounter{subsectionc}{0}\setcounter{subsubsectionc}{0}\noindent 
	{\tenbf\thesectionc. #1}\par\vspace{5pt}}
\renewcommand{\subsection}[1] {\vspace{12pt}\addtocounter{subsectionc}{1} 
	\setcounter{subsubsectionc}{0}\noindent 
	{\bf\thesectionc.\thesubsectionc. {\kern1pt \bfit #1}}\par\vspace{5pt}}
\renewcommand{\subsubsection}[1] {\vspace{12pt}\addtocounter{subsubsectionc}{1}
	\noindent{\tenrm\thesectionc.\thesubsectionc.\thesubsubsectionc.
	{\kern1pt \tenit #1}}\par\vspace{5pt}}
\newcommand{\nonumsection}[1] {\vspace{12pt}\noindent{\tenbf #1}
	\par\vspace{5pt}}

\newcounter{appendixc}
\newcounter{subappendixc}[appendixc]
\newcounter{subsubappendixc}[subappendixc]
\renewcommand{\thesubappendixc}{\Alph{appendixc}.\arabic{subappendixc}}
\renewcommand{\thesubsubappendixc}
	{\Alph{appendixc}.\arabic{subappendixc}.\arabic{subsubappendixc}}

\renewcommand{\appendix}[1] {\vspace{12pt}
        \refstepcounter{appendixc}
        \setcounter{figure}{0}
        \setcounter{table}{0}
        \setcounter{lemma}{0}
        \setcounter{theorem}{0}
        \setcounter{corollary}{0}
        \setcounter{definition}{0}
        \setcounter{equation}{0}
        \renewcommand{\thefigure}{\Alph{appendixc}.\arabic{figure}}
        \renewcommand{\thetable}{\Alph{appendixc}.\arabic{table}}
        \renewcommand{\theappendixc}{\Alph{appendixc}}
        \renewcommand{\thelemma}{\Alph{appendixc}.\arabic{lemma}}
        \renewcommand{\thetheorem}{\Alph{appendixc}.\arabic{theorem}}
        \renewcommand{\thedefinition}{\Alph{appendixc}.\arabic{definition}}
        \renewcommand{\thecorollary}{\Alph{appendixc}.\arabic{corollary}}
        \renewcommand{\theequation}{\Alph{appendixc}.\arabic{equation}}
        \noindent{\tenbf Appendix \theappendixc #1}\par\vspace{5pt}}
\newcommand{\subappendix}[1] {\vspace{12pt}
        \refstepcounter{subappendixc}
        \noindent{\bf Appendix \thesubappendixc. {\kern1pt \bfit #1}}
	\par\vspace{5pt}}
\newcommand{\subsubappendix}[1] {\vspace{12pt}
        \refstepcounter{subsubappendixc}
        \noindent{\rm Appendix \thesubsubappendixc. {\kern1pt \tenit #1}}
	\par\vspace{5pt}}

\newcommand{\textlineskip}{\baselineskip=13pt}
\newcommand{\smalllineskip}{\baselineskip=10pt}

\def\eightcirc{
\begin{picture}(0,0)
\put(4.4,1.8){\circle{6.5}}
\end{picture}}
\def\eightcopyright{\eightcirc\kern2.7pt\hbox{\eightrm c}} 

\newcommand{\copyrightheading}[1]
	{\vspace*{-2.5cm}\smalllineskip{\flushleft
	{\footnotesize International Journal of Modern Physics C, #1}\\
	{\footnotesize $\eightcopyright$\, World Scientific Publishing
	 Company}\\
	 }}

\newcommand{\publisher}[2]{{\begin{center}\footnotesize\smalllineskip 
	Received #1\\
	Revised #2
	\end{center}
	}}

\def\abstracts#1#2#3{{
	\centering{\begin{minipage}{4.5in}\baselineskip=10pt\footnotesize
	\parindent=0pt #1\par 
	\parindent=15pt #2\par
	\parindent=15pt #3
	\end{minipage}}\par}} 

\newcommand{\bibit}{\nineit}
\newcommand{\bibbf}{\ninebf}
\renewenvironment{thebibliography}[1]
        {\frenchspacing
	 \ninerm\baselineskip=11pt
         \begin{list}{\arabic{enumi}.}
        {\usecounter{enumi}\setlength{\parsep}{0pt}     
	 \setlength{\leftmargin 12.7pt}{\rightmargin 0pt} 
         \setlength{\itemsep}{0pt} \settowidth
	{\labelwidth}{#1.}\sloppy}}{\end{list}}

\newcounter{itemlistc}
\newcounter{romanlistc}
\newcounter{alphlistc}
\newcounter{arabiclistc}

\newcommand{\fcaption}[1]{
        \refstepcounter{figure}
        \setbox\@tempboxa = \hbox{\footnotesize Fig.~\thefigure. #1}
        \ifdim \wd\@tempboxa > 5in
           {\begin{center}
        \parbox{5in}{\footnotesize\smalllineskip Fig.~\thefigure. #1}
            \end{center}}
        \else
             {\begin{center}
             {\footnotesize Fig.~\thefigure. #1}
              \end{center}}
        \fi}

\newcommand{\tcaption}[1]{
        \refstepcounter{table}
        \setbox\@tempboxa = \hbox{\footnotesize Table~\thetable. #1}
        \ifdim \wd\@tempboxa > 5in
           {\begin{center}
        \parbox{5in}{\footnotesize\smalllineskip Table~\thetable. #1}
            \end{center}}
        \else
             {\begin{center}
             {\footnotesize Table~\thetable. #1}
              \end{center}}
        \fi}

\def\@citex[#1]#2{\if@filesw\immediate\write\@auxout
	{\string\citation{#2}}\fi
\def\@citea{}\@cite{\@for\@citeb:=#2\do
	{\@citea\def\@citea{,}\@ifundefined
	{b@\@citeb}{{\bf ?}\@warning
	{Citation `\@citeb' on page \thepage \space undefined}}
	{\csname b@\@citeb\endcsname}}}{#1}}

\newif\if@cghi
\def\cite{\@cghitrue\@ifnextchar [{\@tempswatrue
	\@citex}{\@tempswafalse\@citex[]}}
\def\citelow{\@cghifalse\@ifnextchar [{\@tempswatrue
	\@citex}{\@tempswafalse\@citex[]}}
\def\@cite#1#2{{$\null^{#1}$\if@tempswa\typeout
	{IJCGA warning: optional citation argument 
	ignored: `#2'} \fi}}

\def\pmb#1{\setbox0=\hbox{#1}
	\kern-.025em\copy0\kern-\wd0
	\kern.05em\copy0\kern-\wd0
	\kern-.025em\raise.0433em\box0}

\def\fnt#1#2{\footnotetext{\kern-.3em
	{$^{\mbox{\scriptsize #1}}$}{#2}}}

\def\fpage#1{\begingroup
\voffset=.3in
\thispagestyle{empty}\begin{table}[b]\centerline{\footnotesize #1}
	\end{table}\endgroup}

\def\runninghead#1#2{\pagestyle{myheadings}
\markboth{{\protect\footnotesize\it{\quad #1}}\hfill}
{\hfill{\protect\footnotesize\it{#2\quad}}}}
\font\tenrm=cmr10
\font\tenit=cmti10 
\font\tenbf=cmbx10
\font\bfit=cmbxti10 at 10pt
\font\ninerm=cmr9
\font\nineit=cmti9
\font\ninebf=cmbx9
\font\eightrm=cmr8

\def\qed{\hbox{${\vcenter{\vbox{			
   \hrule height 0.4pt\hbox{\vrule width 0.4pt height 6pt
   \kern5pt\vrule width 0.4pt}\hrule height 0.4pt}}}$}}

\renewcommand{\thefootnote}{\fnsymbol{footnote}}	

\def\bsc{{\sc a\kern-6.4pt\sc a\kern-6.4pt\sc a}}  	
\def\bflatex{\bf L\kern-.30em\raise.3ex\hbox{\bsc}\kern-.14em 
T\kern-.1667em\lower.7ex\hbox{E}\kern-.125em X} 

\begin{document}

\runninghead{D. Cai, C. M. Snell, K. M. Beardmore \&
N. Gr{\o}nbech-Jensen}
{Simulation of Phosphorus Implantation Into Silicon}

\normalsize\textlineskip
\thispagestyle{empty}
\setcounter{page}{1}

\copyrightheading{Vol. 0, No. 0 (1998) 000--000}

\vspace*{0.88truein}

\fpage{1}
\centerline{\bf SIMULATION OF PHOSPHORUS IMPLANTATION INTO SILICON}
\vspace*{0.035truein}
\centerline{\bf WITH A SINGLE PARAMETER ELECTRONIC STOPPING POWER MODEL}
\vspace*{0.37truein}
\centerline{\footnotesize DAVID CAI\footnote{Present Address:
Courant Institute of Mathematical Sciences, New York, NY 10012, USA}}
\vspace*{0.015truein}
\centerline{\footnotesize\it Theoretical Division,
Los Alamos National Laboratory, Los Alamos, NM 87545, USA}
\vspace*{0.15truein}
\centerline{\footnotesize CHARLES M. SNELL} 
\vspace*{0.015truein}
\centerline{\footnotesize\it Applied Theoretical and Computational
Physics Division, Los Alamos National Laboratory,}
\baselineskip=10pt
{\centerline{\footnotesize\it Los Alamos, NM 87545, USA}
\vspace*{0.15truein}
\centerline{\footnotesize KEITH M. BEARDMORE and NIELS GR{\O}NBECH-JENSEN} 
\vspace*{0.015truein}
\centerline{\footnotesize\it Theoretical Division,
Los Alamos National Laboratory, Los Alamos, NM 87545, USA}
\vspace*{0.225truein}
\publisher{3 October 1997}{19 March 1998}

\vspace*{0.21truein}
\abstracts{We simulate dopant profiles for phosphorus implantation into silicon
using a new model for electronic stopping power.
In this model,
the electronic stopping power is factorized into a globally averaged
effective charge $Z_{1}^{*}$, and a local charge density dependent electronic
stopping power for a proton.
There is only a single adjustable parameter in the model,
namely the one electron
radius $r_{s}^{0}$ which controls $Z_{1}^{*}$.
By fine tuning this parameter, we
obtain excellent agreement between simulated dopant profiles and the SIMS
data over a wide range of energies for the channeling case. Our work
provides
a further example of implant species, in addition to
boron and arsenic, to verify the validity of the electronic
stopping power model and to illustrate its generality for studies of
physical processes involving electronic stopping.}{}{}



\vspace*{1pt}\textlineskip	
\section{Introduction}		
\vspace*{-0.5pt}
\noindent
Monte Carlo and molecular dynamics
simulations
of ion trajectories in a target material require a good
description of the physics of electronic stopping in the
high energy regime. The issue of the electronic stopping
power is especially important when the target is
crystalline and ions can propagate along preferred channel directions
in the lattice. For this case, electronic stopping becomes a dominant
factor in determining
final stopping ranges of the channeling ions.
As is well known, the classic Lindhard-Scharff-Schiott
theory\cite{LSS} is applicable only
to amorphous materials and underestimates
electronic stopping along
channeling directions. It thus tends to
give an excessively high estimate of the energy threshold below which
electronic stopping effects can safely be neglected when modeling
ion implantation into crystalline materials. For the channeling
case, a good understanding of electronic stopping
is essential because the distribution of stopping ranges of
ions can still be significantly affected by the
electronic stopping power even at relatively low energies.

\newpage
\textheight=7.8truein		
\setcounter{footnote}{0}
\renewcommand{\thefootnote}{\alph{footnote}}

Recently we proposed a model for electronic stopping power for
ion implantation modeling.\cite{Cai} Based on the spirit
of the Brandt-Kitagawa (BK) effective charge theory,\cite{BK} it models the
electronic stopping power for an ion in terms of two factors: (i)
a globally averaged effective charge taking into account effects
of close and distant collisions by target electrons with the ion,
and (ii) a local charge density  dependent electronic stopping power
for a proton. This model was implemented into both molecular
dynamics and Monte Carlo simulations. The results of the dopant
profile simulation for both boron and arsenic implants into
crystalline and amorphous silicon have
demonstrated that our model can successfully capture
the {\it physics}
of electronic stopping in ion implantation over a wide range of
energies.
The model is phenomenologically economical, i.e., it has only
one tuning parameter, namely an averaged one electron radius $r^{0}_{s}$
which controls the effective charge of the ion.
A single numerical value of this parameter was
used for the simulation of both boron and
arsenic implantation.  Good agreement of
dopant profiles with experimental profiles measured
by secondary-ion mass spectroscopy (SIMS) was achieved
for both species with this single $r^{0}_{s}$ numerical value.

We note that the BK theory uses a statistical model for
the partially ionized projectile and does not account for
shell structure. Therefore, it can only provide an
averaged description of electronic stopping power
as a function of the projectile atomic number $Z_{1}$.\cite{Barberan}
The experimentally observed $Z_{1}$ oscillations in
electronic stopping are
a complex phenomenon, attributable to the
electronic shell structures of both the incident ion and
the target atom.\cite{Nastasi,Echenique,Eisen,Briggs,Wilson}
On account of the dependence of the $Z_1$ oscillation on both the ion
and target material, we expect that the
parameter $r^{0}_{s}$ can be tuned to different
numerical values for different combinations of implant
species and substrate material. This fine tuning
can be viewed as a phenomenological procedure to incorporate
the physics of  $Z_{1}$ oscillations.
In the present work, we
will verify this phenomenological approach for
phosphorus implantation
into silicon
and show that our electronic stopping power can successfully
model channeling of phosphorus implants into single-crystal
silicon, thus extending our electronic stopping power model
to the case of phosphorus-on-silicon implants. This further
illustrates the potential wide applicability of the
model in studies of physical processes that involve
electronic stopping.

The paper is organized as follows.
In Sec.~2 we summarize the main features of our
electronic stopping power model and briefly compare it with other
models that have been used in various Monte
Carlo simulations. Atomic units $e = \hbar = m_{e} =1$
are used in the description of our model. In Sec.~3 we describe
the implementation
of our model into a MARLOWE platform,\cite{marlowe} and
into a molecular dynamics (MD) based implant simulator.\cite{reed}
We then present a comparison between our simulation results
and SIMS data
for phosphorus implantation into silicon.
In Sec.~4 we make concluding remarks.

\newpage

\section{The Model}
\noindent
In our model,\cite{Cai} based on
an effective charge scaling argument,
the electronic stopping power of
an ion can be factorized into two components. One is
the effective charge  $Z_{1}^{*}$ of the ion of velocity $v_{1}$,
which can be expressed
as
\begin{equation}
Z_{1}^{*} = Z_{1} \gamma (v_{1},r_{s}^{0}),
\end{equation}
where $Z_{1}$ is the atomic number of the ion and
$\gamma (v_{1}, r_{s}^{0})$ is the fractional effective charge of
the ion. The second is the charge density dependent
electronic stopping power
for a proton $S_{p}(v_{1},r_{s})$. Here $r_{s}$ is the one electron
radius, $r_{s} = [3/\left(4\pi \rho(\bf x) \right)]^{1/3}$, where
$\rho(\bf x)$ is
the charge density of the target. In our treatment $\gamma(v_{1},r_{s}^{0})$
does not depend on the local charge density, instead, it is controlled
by the parameter $r_{s}^{0}$, which is the only adjustable parameter in
the model.

After taking account of the energy loss of the ion in soft, distant collisions
with target electrons
and the energy loss in hard close collisions,
the BK analysis gives the fractional effective charge
\begin{equation}
\gamma(v_{1},r_{s}^{0}) = q(v_{1},r_{s}^{0}) + C [1-q(v_{1},r_{s}^{0})]
\ln\left[1+\left(\frac{4\Lambda(v_{1},r_{s}^{0})}{r_{s}^{0}}\right)^{2}\right].
\end{equation}
Here $C$  is weakly dependent on the target
and has a numerical value of about 1/2. Below, it is set to be
one-half. The ion size parameter $\Lambda(v_1,r_{s}^{0})$ is
\begin{equation}
\Lambda(v_{1},r_{s}^{0}) = \frac{2a_{0} [1-q(v_{1},r_{s}^{0})]^{2/3}}{Z_{1}^{1/3}
[1-(1-q(v_1,r_{s}^{0}))/7]},
\end{equation}
which is used
in the statistical model to describe the partially ionized
projectile.
Here $a_{0} = 0.24\,005$ and the ionization fraction $q(v_{1},r_{s}^{0})$
obeys the following scaling
\begin{equation}
q(v_{1},r_{1}^{0})= 1 -\exp[-0.95(y_{r}(v_{1},r_{s}^{0})-0.07)].
\label{q}
\end{equation}
This scaling was condensed from extensive experimental data
for ions $3\le Z_{1} \le 92$.\cite{ZBL}
 The reduced relative velocity  $y_{r}(v_{1},r_{s}^{0})$ is defined as
\begin{equation}
y_r(v_{1},r_{s}^{0}) = \frac{v_{r}(v_{1},r_{s}^{0})}{v_{B}Z_{1}^{2/3}},
\end{equation}
where $v_{B}$ is the Bohr velocity.
Underlying Eq. (\ref{q}) is the stripping criterion that
the electrons of the ion which have an orbital velocity lower
than the relative velocity between the ion and the electrons
in the medium are stripped off.
Averaging relative velocity over the conduction
electrons leads to:\cite{Kreussler}
\begin{eqnarray}
v_{r}(v_{1},r_{s}^{0}) & = & v_{1} \left(1 + \frac{v_{F}^{2}}{5 v_{1}^2} \right)
\mbox{\hspace{4mm} for $v_{1} \ge v_{F}$} \\
v_{r}(v_{1},r_{s}^{0}) & = & \frac{3v_F}{4}
\left(1 + \frac{2 v_{1}^2}{3 v_{F}^{2}} - \frac{v_{1}^4}{15v_{F}^4}\right)
\mbox{\hspace{4mm} for $v_{1} < v_{F}$}
\end{eqnarray}
In our model,\cite{Cai} the Fermi velocity is related to $r_{s}^{0}$:
$v_{F} = 1/(\alpha r_{s}^{0})$, $\alpha = [4/(9\pi)]^{1/3}$.

The electronic stopping power for a proton in our model uses results
derived from a nonlinear density-functional formalism:\cite{Echenique-solid}
\begin{equation}
S_{p}(v_1,r_{s}) = -\left(\frac{dE}{dx}\right)_{\rm R} G(r_{s}),
\end{equation}
where
\begin{equation}
\left(\frac{dE}{dx}\right)_{\rm R} = \frac{2v_{1}}{3\pi}
\left[ \ln \left(1 + \frac{\pi}{\alpha r_{s}} \right)
- \frac{1}{1+\alpha r_{s}/\pi}\right]
\end{equation}
is the Ritchie formula for the energy loss per unit path
length of a proton moving at velocity $v_1$ in the electron
gas derived from a linear response theory. The
correction factor $G(r_{s})$
is a  computationally convenient way to incorporate
the density-functional results and it has the following
form
\begin{equation}
G(r_{s})
=
1.00 + 0.717 r_{s}
-0.125 r_{s}^{2}
-0.0124 r_{s}^3
+0.00212 r_{s}^4
\end{equation}
for $r_{s} < 6 $. It should be emphasized that
$S_{p}(v_{1},r_{s})$ in our model depends on
the local charge density, $\rho({\bf x}) = 3/(4\pi r_{s}({\bf x})^{3})$.
We note in passing that
the density-functional result gives a better
estimate than the linear response (dielectric)
result for the stopping powers as demonstrated
in the comparisons with experimental data.\cite{Mann,Brandt}
The charge density, $\rho(\bf x)$, for atoms in crystal
silicon in our model uses the solid-state Hartree-Fock atomic
charge distribution.\cite{ZBL} In this approximation,
there is about a one electron charge not accounted for
inside the spherical muffin-tin. This small amount of
charge
is distributed between the maximal collision distance
used in our Monte Carlo simulations and the muffin-tin radius;\cite{Cai}
within the MD scheme it provides a background
charge experienced by the ion when further than the muffin-tin radius
from any silicon atom.\cite{reed}

In our simulation, the electronic stopping power is evaluated continuously
along the path the ion traverses through regions of varying
charge density. The energy loss is computed by
\begin{equation}
\Delta E_{e} = \int_{\rm ion\,\, path} [Z_{1}\gamma(v_{1}, r_{s}^{0})]^{2}
S_{p}(v_{1},r_{s}(x))dx.
\end{equation}

Finally, we discuss the difference between our approach and other
electronic stopping power models used in Monte Carlo simulations
based on the MARLOWE platform. There is a purely nonlocal version
of the BK theory implemented into MARLOWE,\cite{Azziz} where the
effective charge and the stopping power for a proton both depend
on a single nonlocal parameter, i.e., the averaged one electron
radius. The energy loss for well-channeled ions in the keV region in this
approach indicated that a correct density distribution is necessary to
model the electronic stopping in the channel.\cite{Azziz,Murthy}
Later, an implementation of a purely local version of
the BK theory to model boron implants into $\langle 100 \rangle$
single-crystal silicon produced dopant profiles in good agreement
with the SIMS data.\cite{Klein} It
showed a marked improvement in modeling
dopant profiles in the channel over other electronic stopping power
models, such as Lindhard and Scharff,\cite{Scharff} Firsov,\cite{Firsov}
and the above nonlocal implementation.  However,
this purely local version was unable to model the electronic
stopping either for boron implants into the $\langle 110 \rangle $ axial
channel, or for arsenic implants.\cite{Yang}

In our model, the effective
charge is a nonlocal quantity, neither explicitly dependent on the impact
parameter nor on the charge distribution, while the stopping power for
a proton depends on the local charge density of silicon. It has
been shown that it can successfully model both boron and arsenic
implants into silicon over a wide range of energies and in different
channeling and off-axis directions of incidence.\cite{Cai,reed} In the
following section, we will demonstrate that this model can be further
extended to phosphorus implantation into silicon. In light of $Z_{1}$
oscillations, $r_{s}^{0}$ will take a slightly different numerical value
from the one used in the simulations for
boron and arsenic implantation.\cite{Cai}

\section{Monte Carlo and Molecular Dynamics Simulation Results}
\noindent
We have implemented the electronic stopping power model
into the Monte Carlo simulation platform MARLOWE, which utilizes
the binary collision approximation (BCA) to simulate the trajectories
of energetic ions in crystalline or 
amorphous materials.\cite{Robinson,Eckstein}
For the present work, we used an extended version of MARLOWE with
enhanced capabilities for modeling ion implant into silicon.
This code, UT-MARLOWE,\cite{marlowe} was selected for its more versatile
features, including variance reduction for more efficient calculation,
and the incorporation of important implant parameters, e.g., tilt and rotation
angles, the thickness of the native oxide layers, beam divergence, and
wafer temperature.
The electronic stopping power model is also incorporated into an MD based
implant simulator, REED.\cite{reed} This allows us to calculate profiles
using a more realistic description of atomic collisions during implants
than that provided by the BCA. It also gives us an additional verification
of the model, and shows that it is independent of the simulation platform.

For the purpose of
verifying the electronic stopping model,
the option of the cumulative damage model in the UT-MARLOWE code
was turned off in  our simulations.
This is a phenomenological model for estimating
defect production and recombination rates. The REED cumulative damage model
was also disabled.
In order to minimize
cumulative damage effects in the dopant profiles,
simulations were performed for
low to moderate dose ($10^{13}$~cm$^{-2}$ to $3\times10^{13}$~cm$^{-2}$)
phosphorus implants. Individual ion trajectories
were simulated
and the overlapping of the damage
caused by different individual ion cascades was neglected.
Also, use was made of $1^{\circ}$ initial beam divergence,
a 16~\AA\  native oxide surface layer, and 300 K wafer
temperature. These parameters were not cited in the experimental
references but are believed to be typical for silicon doping implants.

The best suited data for verifying the electronic stopping model
are the on-axis or $\langle 100 \rangle$ channeling implants. These data
are highly sensitive to electronic stopping and are
relatively insensitive to other effects, especially at
energies of 100 keV and above where electronic stopping
becomes the dominant energy loss mechanism.
The free parameter in the model $r_{s}^{0}$
was adjusted to 1.148~\AA\ to
yield the best results in overall comparison between the
BCA and SIMS profiles.
A fixed numerical value of $r_{s}^{0}$ was
employed for all energies and directions of incidence. As commented
before,\cite{Cai} this value is physically reasonable for silicon.
We note that there is only a $3.5\%$ difference
between this value of $r_{s}^{0}$
and the value of 1.109~\AA\
previously used to obtain the best overall agreement
with experimental data for boron and arsenic.\cite{Cai}
This difference is quite small, and our result indicates that,
in general, $Z_{1}$ oscillations can be accounted for by
fine-tuning of $r_{s}^{0}$ for a given combination of
implant species and material.
A slightly larger value of 1.217~\AA\ was used for $r_{s}^{0}$
within the MD simulations. This difference reflects the fact that the
description of ion channelling and energy loss differs between the BCA
and MD schemes,
and that while fitting the electronic stopping model we are also compensating
for deficiencies in the implant simulation models.

\begin{figure}[ht!]
\vbox{\vspace{-2.20 in} \hspace{0.0 in}
\includegraphics{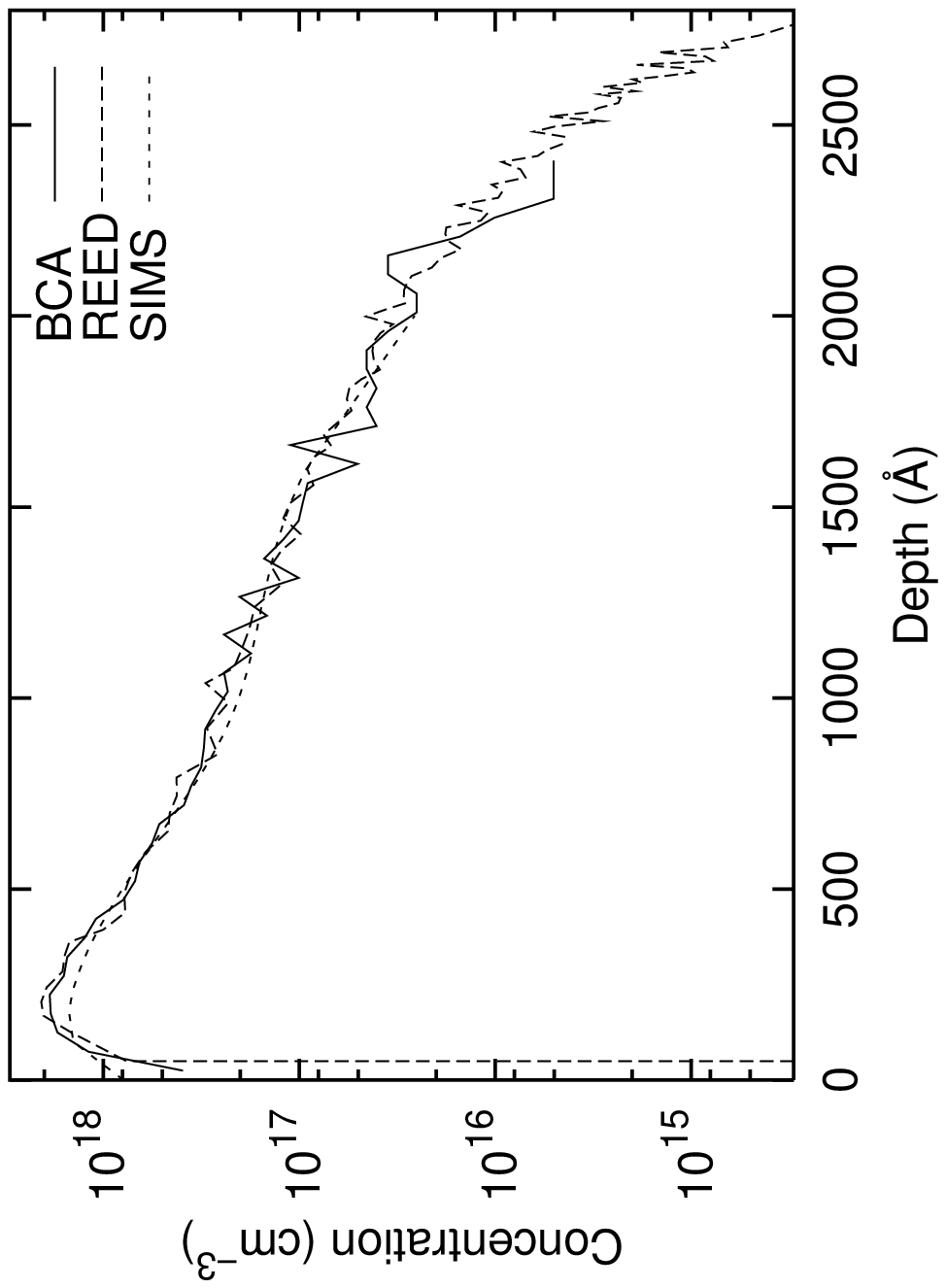}
\vspace{4.40 in}}

\fcaption{Calculated and experimental dopant profiles due to
15 keV phosphorus implant for the $\langle 100 \rangle$
direction with zero tilt and rotation angles;
the dose is $10^{13}$~cm$^{-2}$.
\vspace{-0.22 in} }
\label{15k-00-1}
\end{figure}

\begin{figure}[ht!]
\vbox{\vspace{-2.15 in} \hspace{0.0 in}
\includegraphics{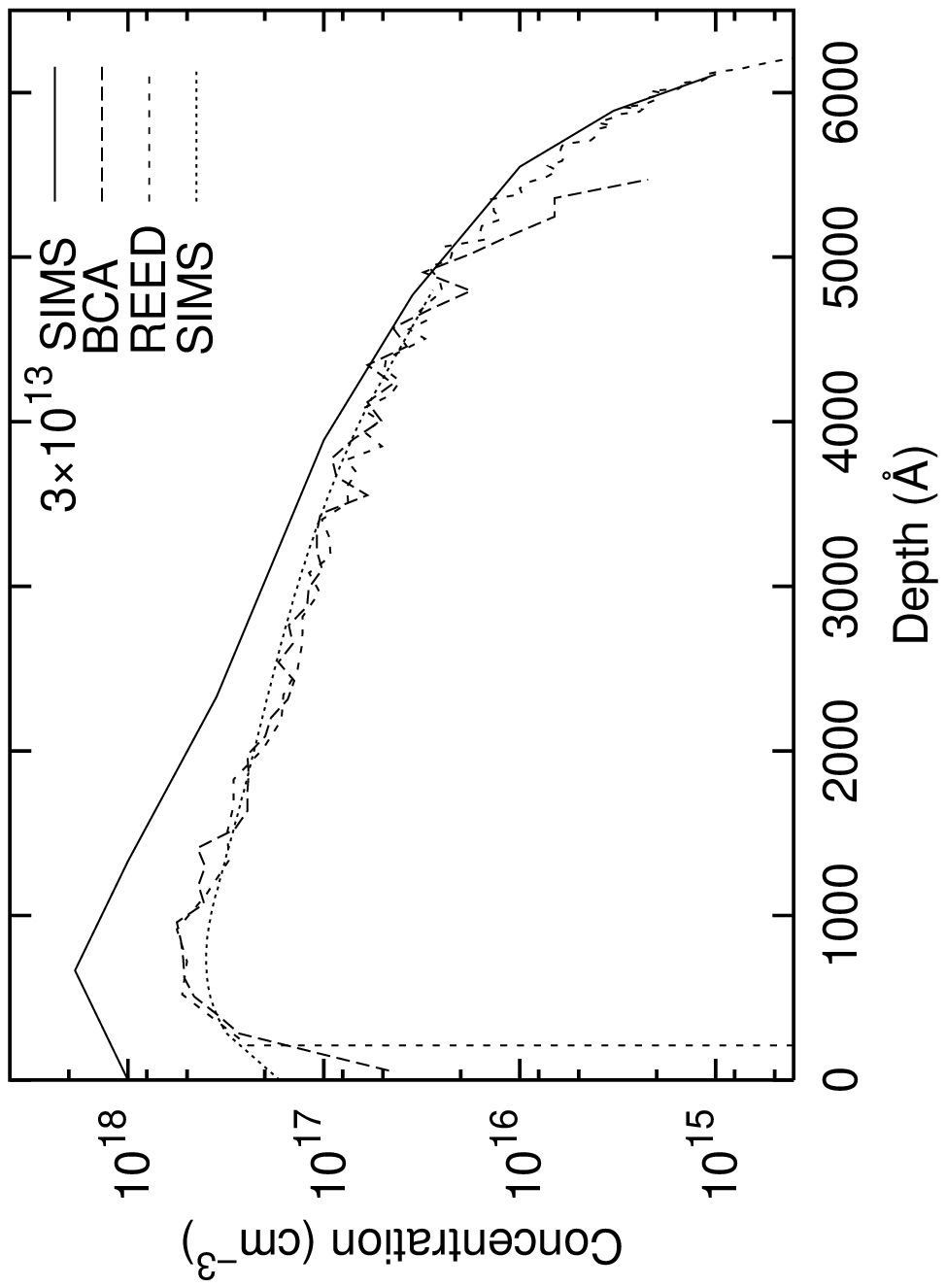}
\vspace{4.50 in}}

\fcaption{Calculated and experimental dopant profiles due to
50 keV phosphorus implant for the $\langle 100 \rangle$
direction with zero tilt and rotation angles;
the dose is $10^{13}$~cm$^{-2}$ -- the SIMS profile for a dose of
$3\times10^{13}$~cm$^{-2}$ is also shown to illustrate damage effects.
\vspace{-0.22 in} }
\label{50k-00-1}
\end{figure}

\begin{figure}[ht!]
\vbox{\vspace{-2.15 in} \hspace{0.0 in}
\includegraphics{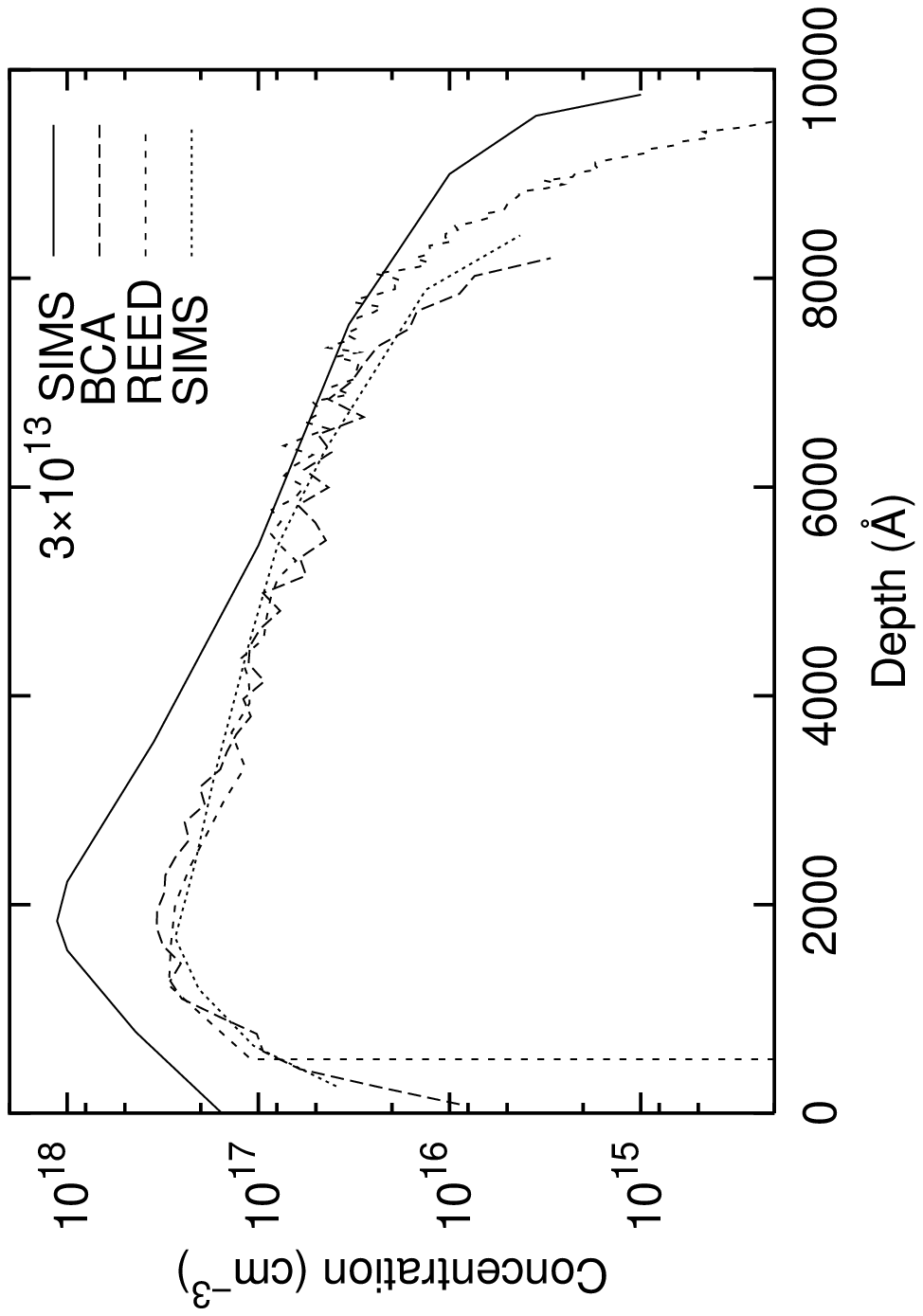}
\vspace{4.45 in}}

\fcaption{Calculated and experimental dopant profiles due to
100 keV phosphorus implant for the $\langle 100 \rangle$
direction with zero tilt and rotation angles;
the dose is $10^{13}$~cm$^{-2}$ -- the SIMS profile for a dose of
$3\times10^{13}$~cm$^{-2}$ is also shown to illustrate damage effects.
\vspace{-0.25 in} }
\label{100k-00-1}
\end{figure}

We now turn to our Monte Carlo and MD dopant profile simulation results.
For comparison we have digitized the
SIMS data for phosphorus reported in
Refs.~9 and~25.
The simulated phosphorus dopant profiles are shown
in comparison to the SIMS data for the channeling case in Figs.
\ref{15k-00-1}, \ref{50k-00-1}, \ref{100k-00-1},
and \ref{200k-00-1}.
Here, the implant energy ranges
from 15 keV
to 200 keV.  The incidence is
along the $\langle 100 \rangle$ direction with
zero tilt and zero rotation angles. The dose is
$10^{13}$~cm$^{-2}$ or $3\times10^{13}$~cm$^{-2}$.
The simulations show a close fit to the dopant distributions with
depth and describe especially well
the slope of the channeling tails of the dopant profiles.
Some uncertainty is introduced into these comparisons by the
fact that the exact implant parameters employed in the experiments
are not known. Ion channeling ranges for on-axis implants may be
sensitive to variations in these parameters, especially to the
beam divergence. Fig. \ref{beam-comp} illustrates the effect of changing
the assumed beam divergence from $1^{\circ}$ to $0^{\circ}$.
As expected, the lower divergence gives a shallower slope
and deeper penetration in the channeling tail. Best agreement
with the SIMS profile is achieved for a divergence angle
of $1^{\circ}$, which is quite reasonable for commercial
implant machines. This value was employed in all other
calculations presented here.

\begin{figure}[ht!]
\vbox{\vspace{-2.15 in} \hspace{0.0 in}
\includegraphics{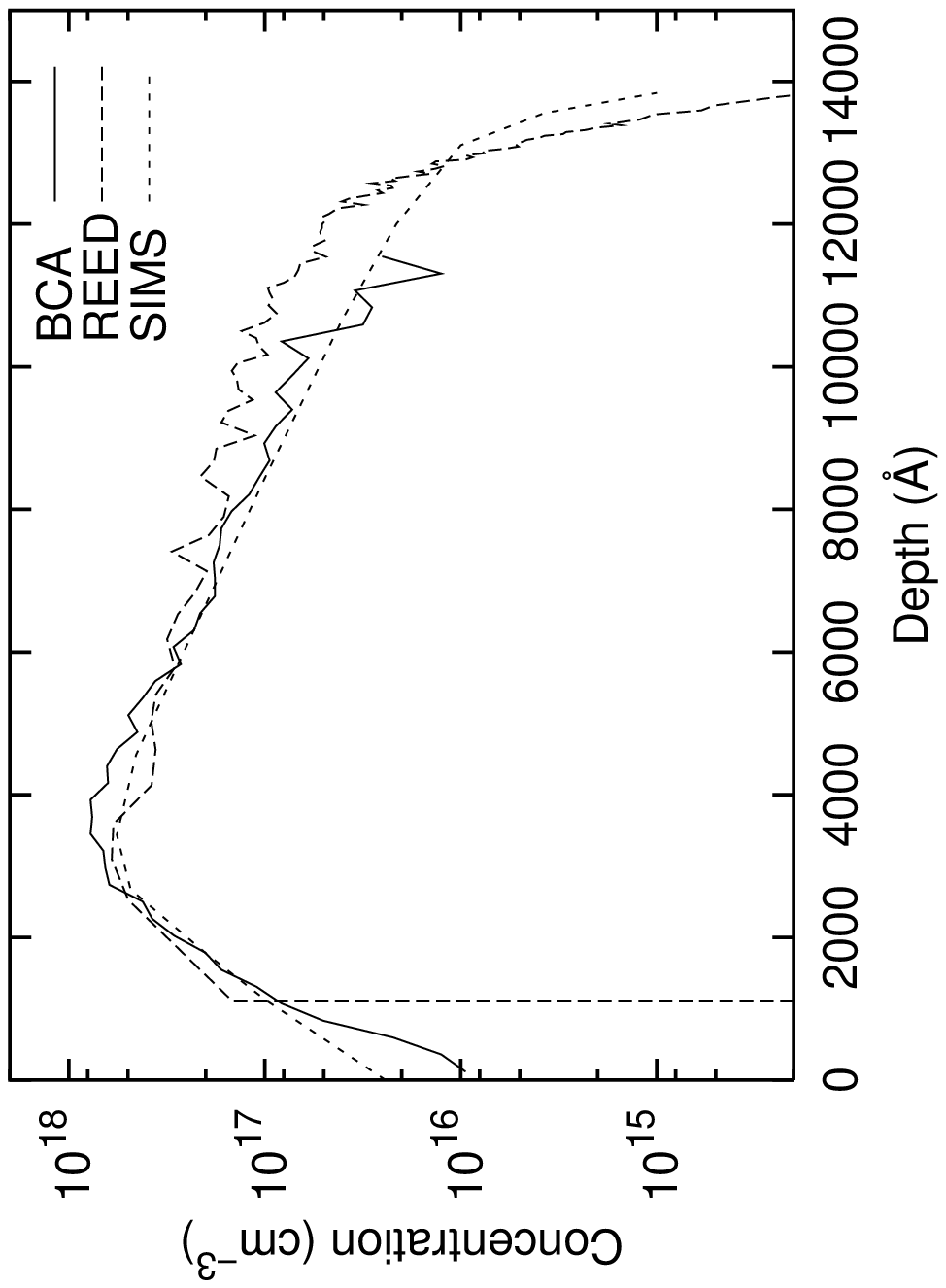}
\vspace{4.40 in}}

\fcaption{Calculated and experimental dopant profiles due to
200 keV phosphorus implant for the $\langle 100 \rangle$
direction with zero tilt and rotation angles;
the dose is $3\times10^{13}$~cm$^{-2}$.
\vspace{-0.22 in} }
\label{200k-00-1}
\end{figure}

\begin{figure}[ht!]
\vbox{\vspace{-2.10 in} \hspace{0.0 in}
\includegraphics{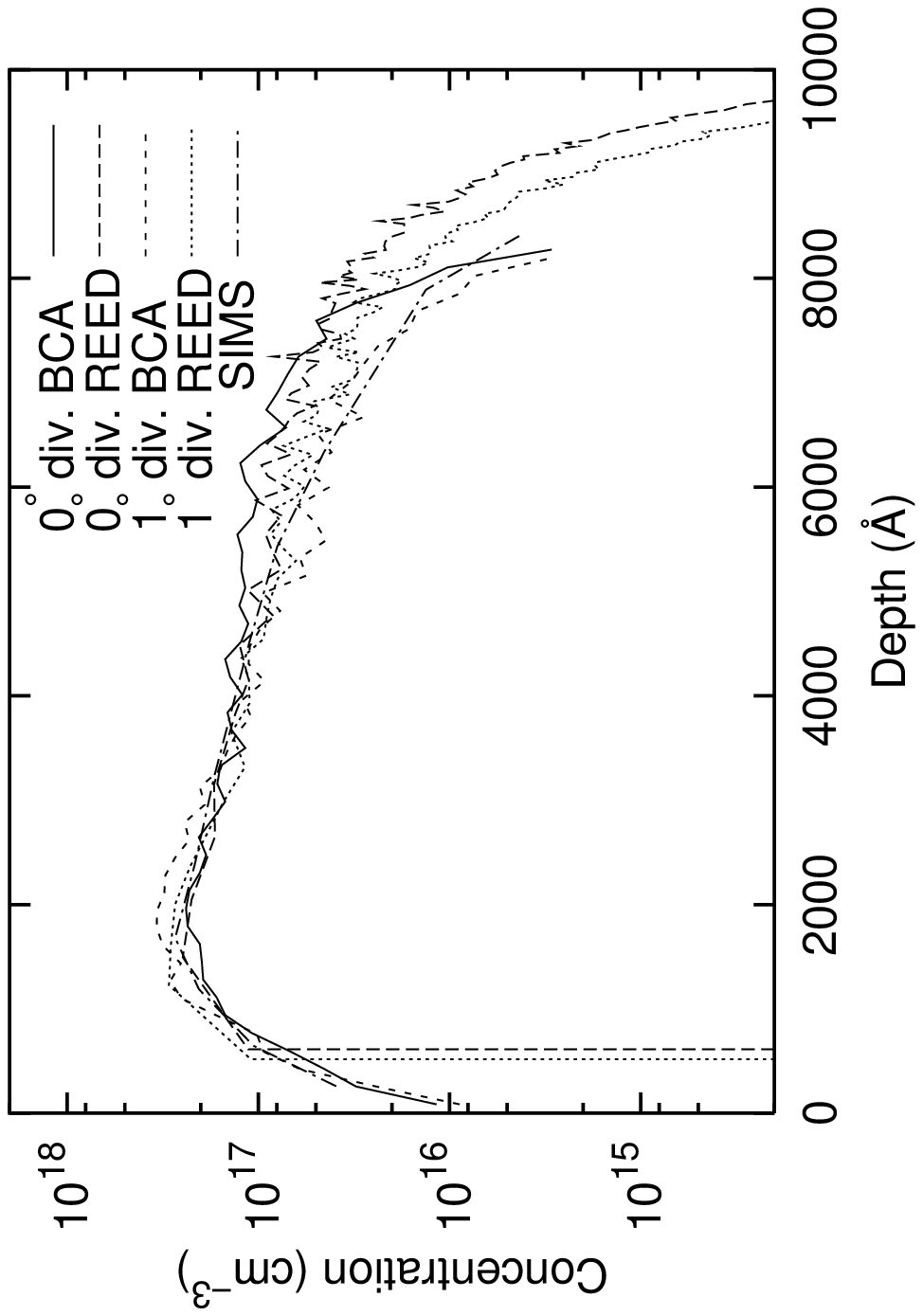}
\vspace{4.40 in}}

\fcaption{Comparison of profiles calculated with $0^{\circ}$ and
$1^{\circ}$ beam divergence; other parameters are as Fig. \ref{100k-00-1}.
\vspace{-0.22 in} }
\label{beam-comp}
\end{figure}

Another source of uncertainty
is the possible influence of disorder introduced by damage to
the silicon, which can reduce the average range of channeled ions.
This effect was ignored in our calculations. The importance of
damage can be roughly evaluated by comparing the measured
SIMS profiles at different doses, as in Figs.~\ref{50k-00-1}
and~\ref{100k-00-1}.
It is seen that the profiles at the higher dose are
similar the lower dose profiles, but that a three-fold increase
in dose results in an almost five-fold increase in dopant concentration
at the peak while only doubling the concentration in the tail.
However the depth of the peak and end-of-range of the profile
are unaltered by the addition of damage. Hence, we can compare the
`zero damage' simulations to the higher dose
($3\times10^{13}$~cm$^{-2}$) SIMS data,
but must be aware that the slope of the channelling tail may
not match -- the MD profile in Fig. \ref{200k-00-1} shows this effect.
We conclude that the overall agreement between the
simulations and experimental data
for the channeling cases is excellent over a wide range of
energies -- especially considering that the electronic stopping model
implemented in the BCA platform has
been tuned by only 3.5\% in the parameter, $r_s^{(0)}$.

Figs. \ref{100k-1015} and \ref{200k-818}
show the simulated dopant profiles in the off-axis directions,
$10^{\circ}$ tilt and $15^{\circ}$ rotation for 100 keV, and
$8^{\circ}$ tilt and $18^{\circ}$ rotation for 200 keV, respectively.
The BCA calculated profiles show less good agreement with
SIMS data than for channeling cases, with
systematically deeper penetration
and the concentration peak shifted by
about $25\%$ -- a similar, but smaller shift is also seen for the
200 keV on-axis implant (Fig. \ref{200k-00-1}).

\begin{figure}[ht!]
\vbox{\vspace{-2.15 in} \hspace{0.0 in}
\includegraphics{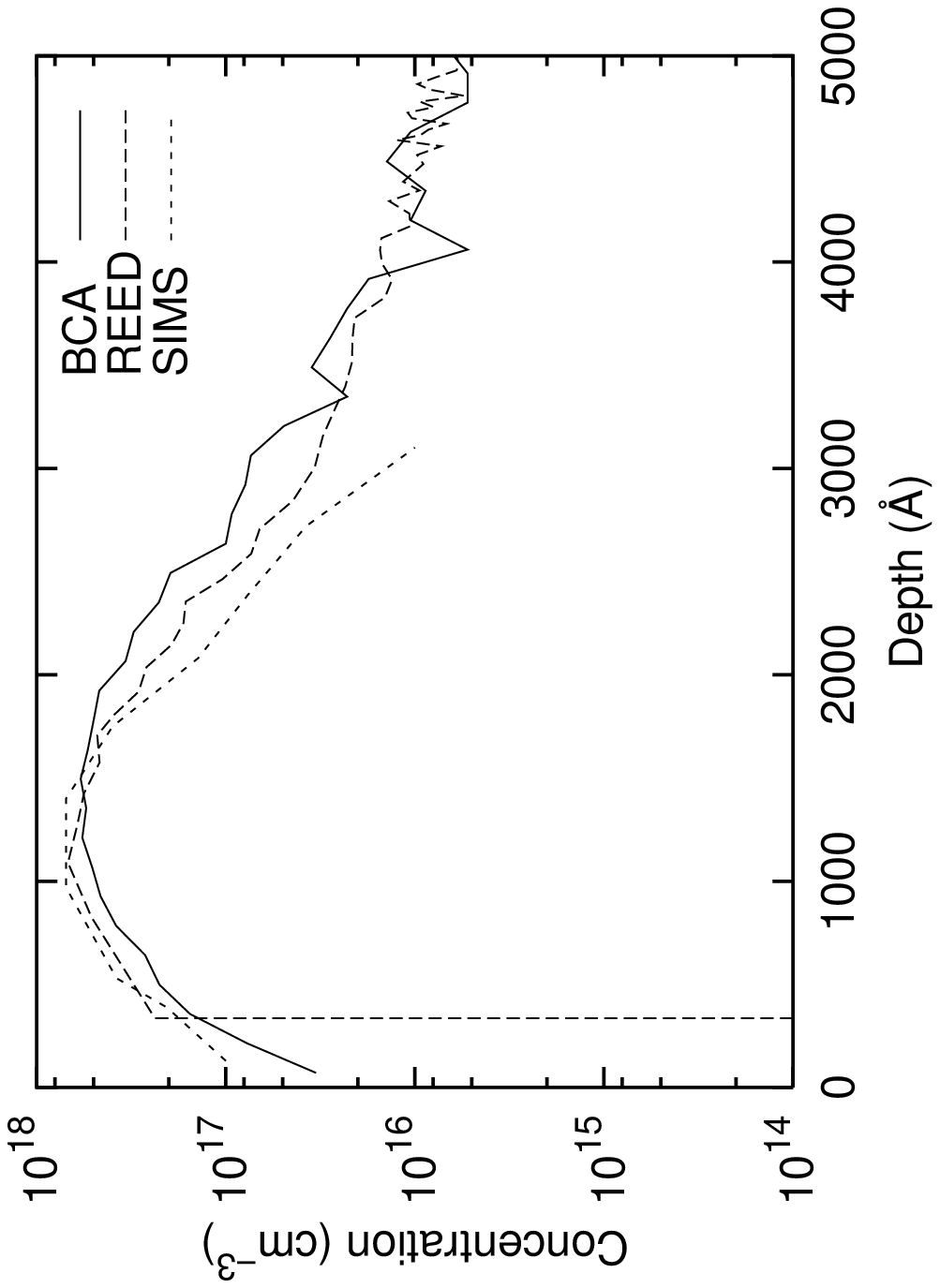}
\vspace{4.40 in}}

\fcaption{Calculated and experimental dopant profiles due to
100 keV phosphorus implant for the $\langle 100 \rangle$
direction with $10^{\circ}$ tilt and $15^{\circ}$ rotation;
the dose is $10^{13}$~cm$^{-2}$.
\vspace{-0.22 in} }
\label{100k-1015}
\end{figure}

\begin{figure}[ht!]
\vbox{\vspace{-2.10 in} \hspace{0.0 in}
\includegraphics{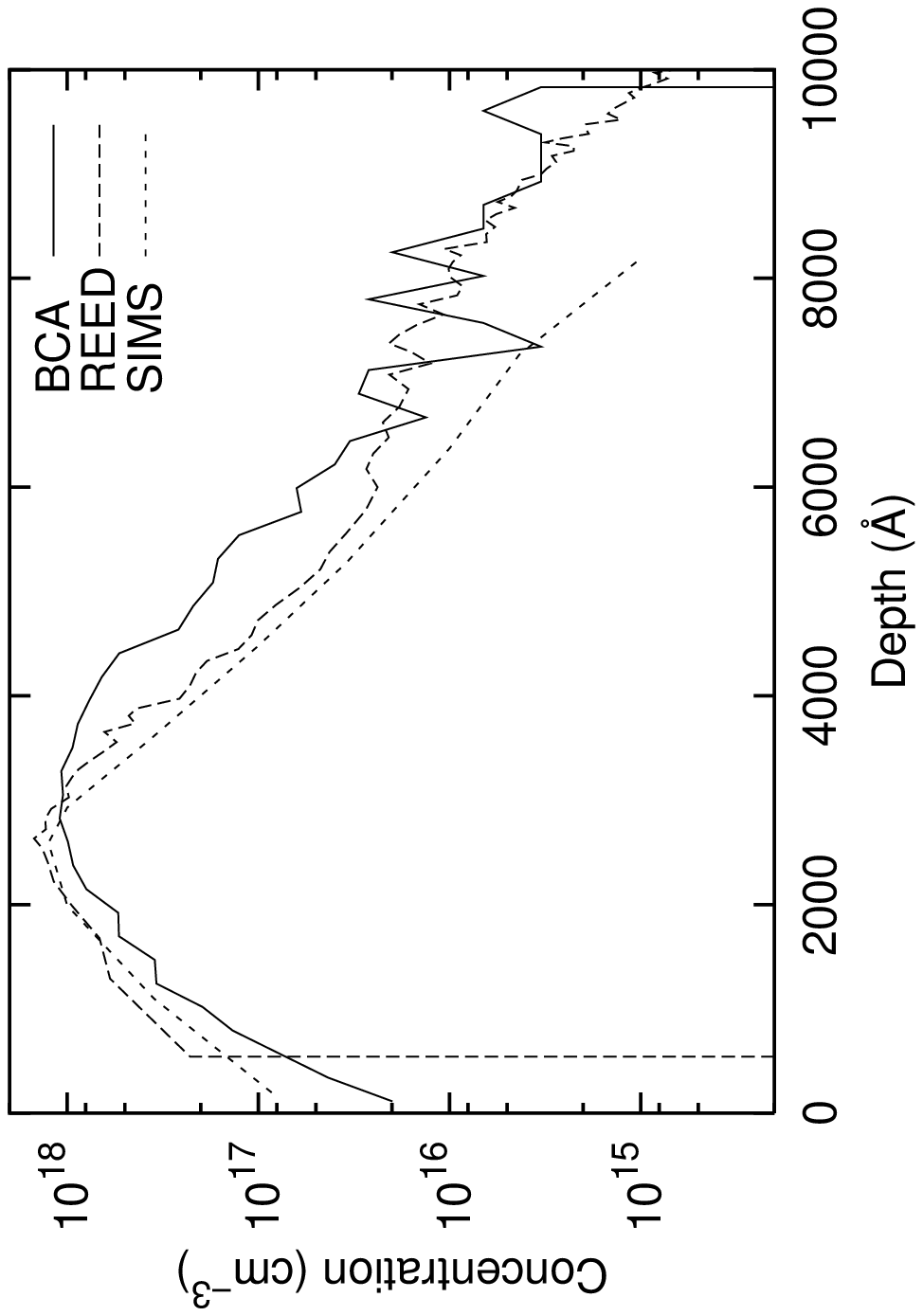}
\vspace{4.40 in}}

\fcaption{Calculated and experimental dopant profiles due to
200 keV phosphorus implant for the $\langle 100 \rangle$
direction with $8^{\circ}$ tilt and $18^{\circ}$ rotation;
the dose is $3\times10^{13}$~cm$^{-2}$.
\vspace{-0.22 in} }
\label{200k-818}
\end{figure}

The explanation for this discrepancy lies in
the different roles of energy loss and scattering mechanisms
for the off-axis implants. The average final stopping range
for ions implanted in off-axis directions
is controlled primarily by nuclear scattering,
and also by inelastic energy loss\cite{Firsov} during ion-atom
interactions; it is less
sensitive to the electronic stopping model (except at very high
energies above those investigated in this study).
Accurate prediction
of the off-axis profiles thus requires an accurate model for the
atom-specific interatomic potential of the two species involved in
the implant, plus a description of inelastic energy loss.
We do not currently have an atom-specific potential model for
phosphorus and silicon in the BCA platform, and there is no attempt
to account for inelastic energy loss due to momentum transfer
between electrons during collisions.
The significance of these omissions in the BCA model
was examined by performing separate calculations for
random implant directions and for an implant into amorphous
silicon (not shown here).
These cases are completely dominated
by nuclear scattering and are insensitive to electronic stopping
in this energy range. The penetration ranges were too large
by about $20\%$, verifying that nuclear scattering
and inelastic energy loss are
largely responsible for the over prediction of ranges seen in
the off-axis calculations.

The MD platform (REED) uses pair-specific potentials\cite{ZBL} for all
ion-silicon interactions, and includes an inelastic energy loss
model.\cite{kis62} The off-axis profiles calculated by MD
have the correct peak position and shallower penetration than the
BCA profiles, and provide a better match to the SIMS data.
The BCA predictions could probably be improved by the relatively simple
addition of the two models mentioned above to the BCA platform,
although that is outside the scope of this paper.
Profiles calculated
by both simulation models show an increased amount of channelling
relative to the SIMS profiles. The channelling tail is due to incident
ions being scattered into channelling directions in near-surface
collisions. The number hard collisions near the surface
due to off-axis implants will produce a significant amount of damage,
even at a dose as low as
$10^{13}$~cm$^{-2}$. Hence we should expect
the `zero damage' calculated profiles to show exaggerated channelling
relative to the experimental data.

\section{Conclusion}
\noindent
We have used our  electronic stopping power model
to simulate dopant profiles for phosphorus implantation into silicon.
To account for $Z_{1}$ oscillations, we have appropriately fine-tuned
the single adjustable parameter
$r_{s}^{0}$ in our model to match the phosphorus-silicon case.
The numerical
value of $r_{s}^{0}$ is slightly greater than the one used for
boron and arsenic simulations. Using this
$r_{s}^{0} = 1.148$~\AA,
our BCA results show excellent agreement between the simulated dopant
profiles and the SIMS data over a wide range of energies for
the channeling case. Less accurate but satisfactory results
are obtained for off-axis implants. Detailed agreement in
the off-axis direction would require additional model
development for ion-silicon interactions.
We have also implemented the stopping model in an MD based simulator,
using a slightly larger value for $r_{s}^{0}$. Using the MD implant model
we achieve excellent agreement with SIMS data for both on-axis, and off-axis
implants.
In summary, we have successfully
extended our electronic stopping power
model to encompass phosphorus implantation into crystalline
silicon. We have also indicated how to incorporate  $Z_{1}$ oscillations
with a simple phenomenological approach. We have provided
a further example of implant species
to  verify validity of the model and to demonstrate its generality for
studies of physical processes involving electronic stopping.

\nonumsection{Acknowledgment}
\noindent
This work was performed at Los Alamos National Laboratory
under the auspices of the U.S. Department of Energy.

\nonumsection{References}

\end{document}